\documentstyle[aps,twocolumn,graphicx,ifthen]{revtex}
\newcommand{\myscalebox}[1]{\scalebox{0.4}[0.45]{#1}}

\newcommand{\myscaleboxc}[1]{\scalebox{0.5}[0.5]{#1}}
\newcommand{\mysection}[1]{\section{#1}}
\begin{document}
\title{Analysis of the
statistical behavior of genetic cluster-exact approximation}

\author{Alexander K. Hartmann\\
{\small  hartmann@theorie.physik.uni-goettingen.de}\\
{\small Institut f\"ur theoretische Physik, Bunsenstr. 9}\\
{\small 37073 G\"ottingen, Germany}\\
{\small Tel. +49-551-399570, Fax. +49-551-399631}}

\date{\today}
\maketitle
\begin{abstract}
The genetic cluster-exact approximation algorithm is an efficient
method to calculate ground states of EA spin glasses. The method can
be used to study ground-state landscapes by calculating many
independent ground states for each realization of the disorder. The
algorithm is analyzed with respect to 
the statistics of the ground states and the valleys of the energy
landscape. Furthermore,
 the distribution inside each valley is evaluated. It is shown that
 the algorithm does not lead to a true $T=0$ thermodynamic
 distribution, i.e. each ground state has not the same frequency of
 occurrence when performing many runs. 
An extension of the technique is outlined, which
 guarantees that each ground states occurs with the same probability. 

{\bf Keywords (PACS-codes)}: Spin glasses and other random models (75.10.Nr), 
Numerical simulation studies (75.40.Mg),
General mathematical systems (02.10.Jf). 
\end{abstract}

\mysection{Introduction}

The finite-dimensional 
Edwards-Anderson spin glass \cite{binder86} is a model for
disordered systems which has attracted much attention over the last decades.
The opinion on its nature, especially for three dimensional
systems,  is still controversial 
\cite{parisi2,mcmillan,bray,fisher,bovier,newman}. Beside
trying to address the problem with the help of analytic calculations
and simulations at finite temperature, it is possible to investigate
the behavior of the model by means of ground-state calculations
\cite{rieger98}. Since obtaining spin-glass ground states is
computationally hard \cite{barahona82}, the study is restricted to
relatively small systems.
Recently a new algorithm, the cluster-exact approximation (CEA) 
\cite{alex2} was presented, which allows in
connection with a special genetic algorithm \cite{pal96} the
calculation of true \cite{alex-stiff} ground states for moderate
system sizes, in three
dimensions up to size $14^3$. By applying this method it is possible
to study the ground-state landscape of systems exhibiting a
$T=0$ degeneracy \cite{alex-sg2}. For a thermodynamical correct
evaluation it is necessary that each
ground state contributes to the results with the same weight, since
all ground states have exactly the same energy. Recently it was shown
\cite{alex-false}, that the
genetic CEA causes a bias on the quantities describing the $T=0$
landscape. The aim of this paper is to analyze the algorithm with
respect to its ground-state statistics. The reasons for the deviation
from the correct behavior are given and an extension of the method is
outlined, which guarantees thermodynamical correct results.

In this work, three-dimensional Edwards-Anderson (EA) $\pm J$
spin glasses  are  investigated. They consist of $N$ spins 
$\sigma_i = \pm 1$, described by the Hamiltonian
\begin{equation}
H \equiv - \sum_{\langle i,j\rangle} J_{ij} \sigma_i \sigma_j
\end{equation}

The sum runs over all pairs of nearest neighbors.
The spins are placed on a three-dimensional (d=3) 
cubic lattice of linear size $L$ with periodic boundary conditions in
all directions.
Systems with quenched disorder of the interactions (bonds)
are considered. Their possible values are $J_{ij}=\pm 1$ with equal
probability. To reduce the fluctuations, a constraint is imposed, so that 
$\sum_{\langle i,j\rangle} J_{ij}=0$.

The article is organized as follows: next a description of
the algorithms is presented. Then it is shown for small systems, that
the method does not result in a thermodynamical correct distribution
of the ground states. In section four, the algorithm and its different
variants are analyzed with respect to the ground-state statistics. 
In the last section a summary is given and an extension of the method
is outlined, which should guarantee thermodynamical correct results.

\mysection{Algorithms}

The algorithm for the calculation bases on a special genetic
algorithm \cite{pal96,michal92} and on cluster-exact approximation  
\cite{alex2}. CEA is  an optimization method designed specially 
for spin glasses. Its basic idea is to transform the spin glass in a
way that graph-theoretical methods can be applied, which work only for systems
exhibiting no bond-frustrations. 
Now a short sketch of these algorithms is given, because later the
influence of different variants on the results is discussed.

Genetic algorithms are biologically motivated. An optimal
solution is found by treating many instances of the problem in
parallel, keeping only better instances and replacing bad ones by new
ones (survival of the fittest).
The genetic algorithm starts with an initial population of $M_i$
randomly initialized spin configurations (= {\em individuals}),
which are linearly arranged using an array. The last one is also neighbor of
the first one. Then $n_o \times M_i$ times two neighbors from the population
are taken (called {\em parents}) and two new configurations called
{\em offspring} are created. For that purpose the {\em triadic crossover}
is used which turned out to be very efficient for spin glasses: 
a mask is used which is a
third randomly chosen (usually distant) member of the population with
a fraction of $0.1$ of its spins reversed. In a first step the
offspring are created as copies of the parents. Then those spins are selected,
 where the orientations of the
first parent and the mask agree \cite{pal95}. 
The values of these spins
are swapped between the two offspring. Then a {\em mutation}
 with a rate of $p_m$
is applied to each offspring, i.e. a randomly chosen fraction  $p_m$ of the
spins is reversed.

Next for both offspring the energy is reduced by applying
CEA:
The method constructs iteratively and randomly 
a non-frustrated cluster of spins.
During the construction of the cluster a local gauge-transformation
of the spin variables is applied so that all interactions between cluster
spins become ferromagnetic.
Fig. \ref{figCEAExample} shows an example of how the construction 
of the cluster works for a small spin-glass system.
To increase the performance, spins adjacent to many unsatisfied bonds 
are more likely to be added to the
cluster. This may introduce a bias on the resulting distribution of
the ground states. Later this scheme (``BIAS'') is compared to a
variant (``SAME''), where all spins may contribute to the cluster with
the same probability.

For 3d $\pm J$ spin glasses each cluster
contains typically 55 percent of all spins.
The  non-cluster spins remain fixed during the following calculation,
they act like local magnetic fields on the cluster spins.
Consequently, the ground state of the gauge-transformed cluster is not trivial,
although all interactions inside the cluster are ferromagnetic.
Since the cluster exhibits no bond-frustration, 
an energetic minimum state for its spins can be  calculated 
in polynomial time by using graph-theoretical methods 
\cite{claibo,knoedel,swamy}: an equivalent network is constructed
\cite{picard1}, the maximum flow is calculated 
\cite{traeff,tarjan} and the spins of the
cluster are set to orientations leading to a minimum in energy. 
Please note, that the ground state of the cluster is degenerate itself,
i.e. the spin
orientations can be chosen in different ways leading all to the same energy. 
It is possible to calculate within one single run  
a special graph, which represents
all ground states of the cluster \cite{picard2}, and select one ground
state randomly. This procedure is called ``BROAD'' here.
On the other hand, one can always choose a certain
ground state of the cluster
directly\footnote{This ground state has the  maximum possible magnetization of
the {\em gauge-transformed} spins among all cluster ground states.}. 
Usually this variant, which is called ``QUICK'' here, is
applied,  because it avoids the
construction of the  special graph. But this again introduces a certain bias
on the resulting distribution of the ground states. Later the
influence of the different methods of choosing ground states is discussed.

This CEA minimization step
is performed $n_{\min}$ times for each offspring.
Afterwards each offspring is compared with one of its parents. The
offspring/parent pairs are chosen in the way that the sum of the phenotypic differences
between them is minimal. The phenotypic difference is defined here as the
number of spins where the two configurations differ. Each
parent is replaced if its energy is not lower (i.e. not better) than the 
corresponding offspring.
After this whole step is conducted $n_o \times M_i$ times, the population
is halved: From each pair of neighbors the configuration 
 which has the higher energy is eliminated. If more than 4
individuals remain the process is continued otherwise it
is stopped and the best individual
is taken as result of the calculation.

The following representation summarizes the algorithm. 

\newlength{\mpwidth}
\setlength{\mpwidth}{\textwidth}
\addtolength{\mpwidth}{-2cm}

\begin{center}

\begin{minipage}[b]{\mpwidth}
\newlength{\tablen}
\settowidth{\tablen}{xxx}
\newcommand{\tabspace}{\hspace*{\tablen}}
\begin{tabbing}
\tabspace \= \tabspace \= \tabspace \= \tabspace \= \tabspace \=
\tabspace \= \kill
{\bf algorithm} genetic CEA($\{J_{ij}\}$,
$M_i$, $n_o$, $p_m$, $n_{\min}$)\\
{\bf begin}\\
\> create $M_i$ configurations randomly\\
\> {\bf while} ($M_i > 4$) {\bf do}\\
\> {\bf begin}\\
\> \> {\bf for} $i=1$ {\bf to} $n_o \times M_i$ {\bf do}\\
\>\> {\bf begin}\\
\>\>\> select two neighbors \\
\>\>\> create two offspring using triadic crossover\\
\>\>\> do mutations with rate $p_m$\\
\>\>\> {\bf for} both offspring {\bf do}\\
\>\>\> {\bf begin}\\
\>\>\>\> {\bf for} $j=1$ {\bf to} $n_{\min}$ {\bf do}\\
\>\>\>\> {\bf begin}\\
\>\>\>\>\> construct unfrustrated cluster of spins\\
\>\>\>\>\> construct equivalent network\\
\>\>\>\>\> calculate maximum flow\\
\>\>\>\>\> construct minimum cut\\
\>\>\>\>\> set new orientations of cluster spins\\
\>\>\>\> {\bf end}\\
\>\>\>\> {\bf if} offspring is not worse than related parent \\
\>\>\>\> {\bf then}\\
\>\>\>\>\> replace parent with offspring\\
\>\>\> {\bf end}\\
\>\> {\bf end}\\
\>\> half population; $M_i=M_i/2$\\
\> {\bf end}\\
\> {\bf return} one configuration with lowest energy\\
{\bf end}
\end{tabbing}

\end{minipage}
\end{center}

The whole algorithm is performed $n_R$ times and all configurations
which exhibit the lowest energy are stored, resulting in $n_G$ statistically
independent ground-state configurations ({\em replicas}). 
A priori nothing about the
distribution of ground states raised by the algorithm is known. Thus,
it may be possible that for one given realization of the disorder
some ground states are more likely to be
returned by the procedure than others. Consequently, any
quantities which are calculated by averaging over many independent
ground states, like the distribution of overlaps, may depend on a bias
introduced by the algorithm. For a thermodynamical correct evaluation all
ground states have to contribute with the same weight, since they all have 
exactly the same energy.

For the preceding work, the distribution of the ground states
determined by the algorithm was taken. 
 The method was utilized to examine the ground state landscape of
two-dimensional \cite{alex-2d} and three-dimensional 
\cite{alex-sg2,alex-ultra} $\pm J$ spin glasses by calculating a small
number of ground states per realization. Some of these results depend 
on the statistics of the ground states, as it will be shown in the next
section for the $d=3$ case.

On the other hand, the main findings of the following investigations are not
affected by the bias introduced by genetic CEA:
the existence of a spin-glass phase for nonzero 
temperature was confirmed for the
three-dimensional spin glass \cite{alex-stiff}. The method
 was applied also to the  $\pm J$ random-bond model to investigate its
$T=0$ ferromagnetic to spin-glass transition\cite{alex-threshold}.
Finally, for small sizes up to $L=8$ all
ground-state valleys were obtained by calculating a huge number of
ground  states per realization and applying a new method called
{\em ballistic search} 
\cite{alex-valleys}.

\mysection{Numerical evidence}

In this section results describing the ground-state landscape of small
three-dimensional $\pm J$ spin glasses are evaluated. It is shown that
the data emerging from the use of raw genetic CEA and from a
thermodynamically correct treatment differ substantially.

Several ground states for small systems of size  $N=L^3=3^3, 4^3, 5^3$
were calculated. 1000 realizations  of the
disorder for $L=3,4$ and 100 realizations for $L=5$ were considered.
The parameters ($M_i$,  $n_o$, $p_m$, $n_{\min}$), for which true
ground states are obtained, are shown in \cite{alex-stiff}.
For all calculations the variants BIAS and QUICK were used to obtain
maximum performance. The effect of different variants on the results
is discussed in the next section.

Two schemes of calculation were applied:

\begin{itemize}
\item[A] For each realization $n_R=40$ runs of genetic CEA 
were performed and all states
  exhibiting the ground-state energy stored. Consequently, this scheme
  reflects the ground-state statistics which is determined solely by 
  the genetic CEA
  method. Configurations which have a higher probability of occurrence
  contribute with a larger weight to the results.
\item[B] For each realization the algorithm was run up to $10^5$
  times. Each particular state was stored only once. For later
  analysis the number of times each state occurred 
  was recorded.  Additionally, a systematic local search was applied to
  add possibly
  missing ground states which are
  related  by flips of free spins to  states already found.
  Finally, a $L=3$ realization exhibits 25 different ground states on average.
  For a $L=4$ realization on average 
  240 states were found and 6900 states for $L=5$.

  For the evaluation of physical quantities every ground state is taken
  with the same probability in this scheme. 
  Thus, the statistics obtained in this way
  reflect the true $T=0$ thermodynamic behavior.
\end{itemize}

To analyze the ground-state landscape, the distribution of overlaps is
evaluated.
For a fixed realization $J=\{J_{ij}\}$ of the exchange interactions and two
replicas
$\{\sigma^{\alpha}_i\}, \{\sigma^{\beta}_i\}$, the overlap \cite{parisi2}
is defined as

\begin{equation}
q^{\alpha\beta} \equiv \frac{1}{N} \sum_i \sigma^{\alpha}_i \sigma^{\beta}_i 
\end{equation}

The ground state of a given realization is characterized by the probability
density $P_J(q)$. Averaging over the realizations $J$, denoted
by $[\,\cdot\,]_{J}$, results in ($Z$ = number of realizations)

\begin{equation}
P(q) \equiv [P_J(q)]_{J} = \frac{1}{Z} \sum_{J} P_J(q) \label{def_P_q}
\end{equation}

Because no external field is present the densities are symmetric:
$P_J(q) = P_J(-q)$ and $P(q) = P(-q)$. So only $P(|q|)$ is relevant.

The result of $P(|q|)$ for $L=5$ is shown in
Fig. \ref{figPqLfive}. For the true thermodynamic result small
overlaps occur less frequent than for the data obtained by the
application of pure genetic CEA. Large overlap
values occur more often. This deviation has an influence on the
way the spin glass behavior is
interpreted. The main controversy about finite-dimensional spin
glasses mentioned at the beginning 
is about the question whether  for the infinite system $P(|q|)$
shows a long tail down to $q=0$ or not 
\cite{parisi2,mcmillan,bray,fisher,bovier,newman}.

To investigate the finite size behavior of $P(|q|)$ the fraction
$X_{0.5}$ of the distribution below $q_0=0.5$ is integrated:

\begin{equation}
X_{q_0} \equiv \int_0^{q_0} P(|q|) \, dq
\end{equation}

The development of $X_{0.5}$ as a function of system size $L$ is shown
in Fig. \ref{figXL}. The datapoints for the larger sizes $L\ge 6$,
obtained using pure genetic CEA, are taken from former calculations
\cite{alex-sg2}. These values are more or less independent
of the system size, while the correct thermodynamic behavior
shows a systematic decrease. Whether for $L\to\infty$ the long tail of
$P(|q|)$ persists cannot be concluded from the data, because the
systems are too small. Nevertheless, the true $T=0$ behavior differs
significantly from the former results.

\mysection{Analysis of genetic CEA}

To understand, why genetic CEA fails in producing the thermodynamical
correct results, in this section the statistics of the ground states,
which is determined by the algorithm, is analyzed directly.

For the case where all ground states were calculated using a huge
number of runs, the frequencies each ground
state occurred were recorded. In Fig. \ref{figHistogramm} the result
for one sample realization of $N=5^3$ is shown. The system has 56 different
ground states. For each state the number of times it was returned by
the algorithm in $10^5$ runs is displayed. 
Obviously the large deviations from state
to state cannot be explained by the presence of statistical
fluctuations. Thus, genetic CEA samples different ground states from
the same realization with different weights.

To make this statement more precise, the following analysis was
performed: 
Two ground states are called {\em neighbors}, if they differ only by the
orientation of one spin. All ground states which are accessible from
each other through this neighbor-relation are defined to be in the
same ground-state {\em valley}.
That means, two ground states belong to the same valley,
if it is possible to move from one state to the other by
flipping only free spins, i.e. without changing the energy.
For all realizations the valleys were determined using a method
presented in \cite{alex-valleys}, which allows to treat systems
efficiently  exhibiting a huge number of ground states.
Then the frequencies $h_V$ for each valley $V$ were computed as the sum of
all frequencies of the states belonging to $V$. In
Fig. \ref{figHistoSample} the result is shown for a sample $N=5^3$
realization, which has 15 different ground state valleys. Large
valleys are returned by the algorithm more frequently, but $h_V$ seems
to grow slower than linearly. A strict linear behavior should hold for
an algorithm which guarantees the correct $T=0$ behavior.

For averaging $h_V$ has to be normalized, because the absolute
values of the frequency differ strongly from realization to
realization, even if the size $|V|$ of a valley, i.e. the number of
ground states belonging to it, is the same. For each realization, 
the normalized frequency
$h^{*}_V$ is measured relatively to the average frequency
$\overline{h}_1$ of all
valleys of size 1: $h^{*}_V\equiv h_V/\overline{h}_1$

If a realization does not exhibit a valley
consisting only of one ground state, the frequency $h_{V_s}$ of the
smallest valley $V_s$ is taken. It is assumed, that the normalized
frequency  exhibits a $h^{*}_V=|V|^{\alpha}$ dependence, which is
justified by the results shown later. Consequently, for the case the
size $|V_s|$ of the smallest valley is larger than one, 
$\overline{h}_1\equiv h_{V_s}/|V_s|^{\alpha}$ is chosen. 
The value of $\alpha$ is determined self-consistently.

The result for $L=3$ of $h^{*}_V$ as a function of the valley-size $|V|$
is presented in Fig. \ref{figHistoLthree}. A value of $\alpha=0.854(3)$
was determined. Please note, that the fluctuations for larger valleys
are higher, because quite often only one valley was available for a
given valley-size. The algebraic form is clearly visible, proving that
genetic CEA overestimates systematically the importance of small ground-state
valleys.

For $L=4$ a value of $\alpha=0.705(3)$ was obtained, while the $L=5$
case resulted in $\alpha=0.642(5)$. Consequently, with increasing
system size, the algorithm fails more and more to sample configurations from
different ground-state valleys according to the size of the valleys. This
explains, why the difference of $X_{0.5}(L)$ between the correct result and
the values obtained in \cite{alex-sg2} increases with growing system size.

Similar results were obtained for two-dimensional systems. For $L=5$ a
self-consistent value of $\alpha=0.650(1)$ was found, while the
treatment of $L=7$ systems resulted in $\alpha=0.659(2)$. Here only a
slight finite-size dependence occurs. This may explain the fact, that
the width of the distribution of overlaps, even calculated only by the
application of pure genetic CEA, seems to scale to zero
\cite{alex-2d}.

In the second section of this paper 
two variants of the algorithm were presented,
which may be able to calculate ground states more equally distributed. To
investigate this issue, similar ground-state calculations were
conducted for $L=4$ and again $h^{*}_V$ was calculated. For the case, were
SAME was used instead of BIAS, a value $\alpha= 0.801(2) $ was
determined self-consistently. Using BROAD instead of QUICK resulted in
$\alpha=0.749(3)$. Finally, by applying SAME and BROAD together, 
$\alpha=0.843(3)$ was obtained. Consequently, applying different
variants of the method decreases the tendency of overestimating small
valleys, but the correct thermodynamic
behavior is not obtained as well. 
Even worse, BROAD and SAME are considerably slower than the
combination of QUICK and BIAS.

So far it was shown, that genetic CEA fails in sampling ground states
from different valleys according the size of the valleys. Now we turn
to the question, whether at least states belonging to the same valley are
calculated with the correct thermodynamic distribution. By
investigating the frequencies of different ground states belonging to
the same valley it was found again, that these configurations are not
equally distributed. But it is possible to study this issue in a more
physical way. For that purpose 
ground states of 100 $L=10$ realizations were
calculated. Then the valley structure was analyzed.
The average distribution of overlaps was evaluated, but only
contributions of pairs of states belonging to the same valley were
considered. For comparison, for the same realizations a long $T=0$
Monte-Carlo (MC) simulation was performed, i.e. 
randomly spins were selected and flipped if they were
free. The ground states were used as starting configurations.
Since a MC simulation ensures the correct thermodynamic
distribution of the states, all ground states of a valley appear with
the same frequency, if the simulation is only long enough. A length of
40 Monte-Carlo steps per spin were found to be sufficient for $L=10$.
The result for the distribution of overlaps $P_{\mbox{\small valley}}(|q|)$
restricted to the valleys is displayed in
Fig. \ref{figPqValley}. Significant differences between the datapoints
from the pure genetic CEA and the correct $T=0$ behavior are visible. 
Consequently, the algorithm does not sample configurations belonging to the
same ground-state valley  with the same weight as well.

\mysection{Conclusion}

In this work the genetic cluster-exact approximation method is
analyzed. The algorithm can be used to calculate many independent true
ground states of EA spin glasses. 
The results from the raw application of the method and from
calculations of {\em all} ground states for small system sizes were
compared. By evaluating the distribution of overlaps is was shown, that
genetic CEA imposes some bias on the
ground-state statistics.  Consequently, the results from the
application of the raw method
do not represent the true $T=0$ thermodynamics. 

To elucidate the behavior of the algorithm the statistics of the
ground states were evaluated directly. It was shown, that different 
ground states have  dissimilar probabilities of occurrence. To understand this
effect better, the ground-state valleys were determined. The genetic CEA
method finds configurations from small ground-state valleys relative
to the size of the valley more often than configurations from large
valleys. Additionally, within a valley the states are not sampled with
the same weight as well. It was shown that two variants of the
algorithm, which decrease its efficiency, weaken the effect, but it
still persists.

Summarizing, two effects are responsible for the biased ground-state
sampling of genetic CEA: small valleys are sampled too frequently and
the distribution within the valleys is not flat. 

For small system sizes it is possible to calculate all ground states,
so one can obtain the true thermodynamic average
directly. But already for $L=5$ there are realizations exhibiting more
than $10^5$ different ground states. Since the ground-state degeneracy
grows exponentially with system size \cite{alex-valleys} larger systems
cannot be treated in this way.
The following receipt should overcome these problems and should allow to
obtain the true thermodynamic $T=0$ behavior for larger systems:

\begin{itemize}
\item Calculate several ground states of a realization using genetic
  CEA.
\item Identify the ground states which belong to the same valleys.
\item Estimate the size of each valley. This can be done using a
  variant of ballistic search \cite{alex-valleys}, 
  which works by flipping free spins
  sequentially, each spin at most once. The number of spins flipped is
  a quite accurate measure for the size of a valley.
\item Sample from each valley a number of ground states, which is
  proportional to the size of the valley. This guarantees, that each
  valley contributes with its proper weight. Each state is obtained
  by performing a $T=0$ MC simulation of sufficient length, starting
  with true ground-state configurations. 
  Since MC
  simulations achieve a thermodynamical correct distribution,
  it is guaranteed that the states within each valley are equally distributed.
\end{itemize}

Please note, that it is not necessary to calculate all ground states
to obtain the true thermodynamic behavior, because it is possible to
estimate the size of a valley by analyzing only some sample ground
states belonging to it. Furthermore, it is even only necessary
to have configurations from the largest valleys available, since they
dominate the ground-state behavior. This condition is fulfilled
 by genetic CEA,
because large valleys are sampled more often than small
valleys, even if small valleys appear  too often relatively.

From the results presented here it is not possible to deduce the
correct $T=0$ behavior of the infinite system, because the system
sizes are too small. Using the scheme outlined above, it is
possible to treat system sizes up to $L=14$ \cite{alex-equi}.

\mysection{Acknowledgements}

The author thanks K. Battacharya and A.W. Sandvik for interesting discussions.
The work was supported by the Graduiertenkolleg
``Modellierung und Wissenschaftliches Rechnen in 
Mathematik und Naturwissenschaften'' at the
{\em In\-ter\-diszi\-pli\-n\"a\-res Zentrum f\"ur Wissenschaftliches
  Rechnen} in Heidelberg and the
{\em Paderborn Center for Parallel Computing}
 by the allocation of computer time. The author announces financial
 support from the DFG ({\em Deutsche Forschungsgemeinschaft}).



\newcommand{\captionCEAExample}
{Example of the cluster-exact approximation method. A part of a spin glass
is shown. The circles represent lattice sites/spins. Straight lines represent
ferromagnetic bonds the jagged lines antiferromagnetic interactions. The
top part shows the initial situation. 
The construction starts with the spin at the center. The bottom part 
displays the final stage.
The spins which belong to the cluster carry a plus or minus sign which
indicates how each spin is transformed, so that only ferromagnetic
interactions remain inside the cluster. All other spins cannot be added
to the cluster because it is not possible to multiply them by $\pm 1$
to turn all adjacent bonds ferromagnetic. 
Please note that many other combinations
of spins can be used to build a cluster without frustration.
}

\newcommand{\captionPqLfive}
{Average distribution of overlaps $P(|q|)$ for $L=5$. The dashed line
  shows the old result obtained by computing about $40$ independent
 ground states per realization
  using genetic CEA. The solid line shows the
  same quantity for the case, where all existing ground states were used for
  the evaluation, i.e. where the correct $T=0$ thermodynamic behavior
  is ensured.
}

\newcommand{\captionXL}
{Average fraction $X_{0.5}$ of the distribution of overlaps for
  $|q|<0.5$ as a function of system size $L$. The upper points (circles)
  where obtained by calculating about $40$ independent
 ground states per realization using genetic CEA. The lower points
 (triangles) show the result ($L=3,4,5$) for the case, 
 where all existing ground states were used for
  the evaluation, i.e. where the correct $T=0$ thermodynamic behavior
  is ensured.
}

\newcommand{\captionHistogramm}
{Number of times each ground state is calculated in $10^5$ runs of
genetic CEA for a
sample realization of $N=5^3$. The realization exhibits 56
ground states, which have significant
different probabilities of being calculated.}

\newcommand{\captionHistoSample}
{Number of times a ground state belonging to a specific valley 
is calculated in $10^5$ runs
  of genetic CEA. The result is shown as a function of the valley
  size $|V|$ and  for one $L=5$ sample realization.
}

\newcommand{\captionHistoLthree}
{Normalized number of times a specific valley $V$ is found by genetic
  CEA for $L=3$. The frequency is normalized so that $h^{*}_V=1$ (see
  text). The probability that a cluster is found increases with the
  size of the cluster, but slower than linearly. The line shows a fit 
$h^{*}_V=|V|^{\alpha}$ with $\alpha=0.854(3)$
}

\newcommand{\captionPqValley}
{Distribution $P_{\mbox{\small valley}}(|q|)$ of overlaps for $L=10$
  restricted
  to pairs of ground states belonging to the same valley. The full
  line shows the result for the case, where the statistics of the
  ground state is determined by the genetic CEA algorithm. The data
  represented by the dashed line was obtained using states which are
  equally distributed within each valley, which was guaranteed by
  performing a $T=0$ MC simulation.
}

\begin{figure}[htb]
\begin{center}
\myscaleboxc{\includegraphics{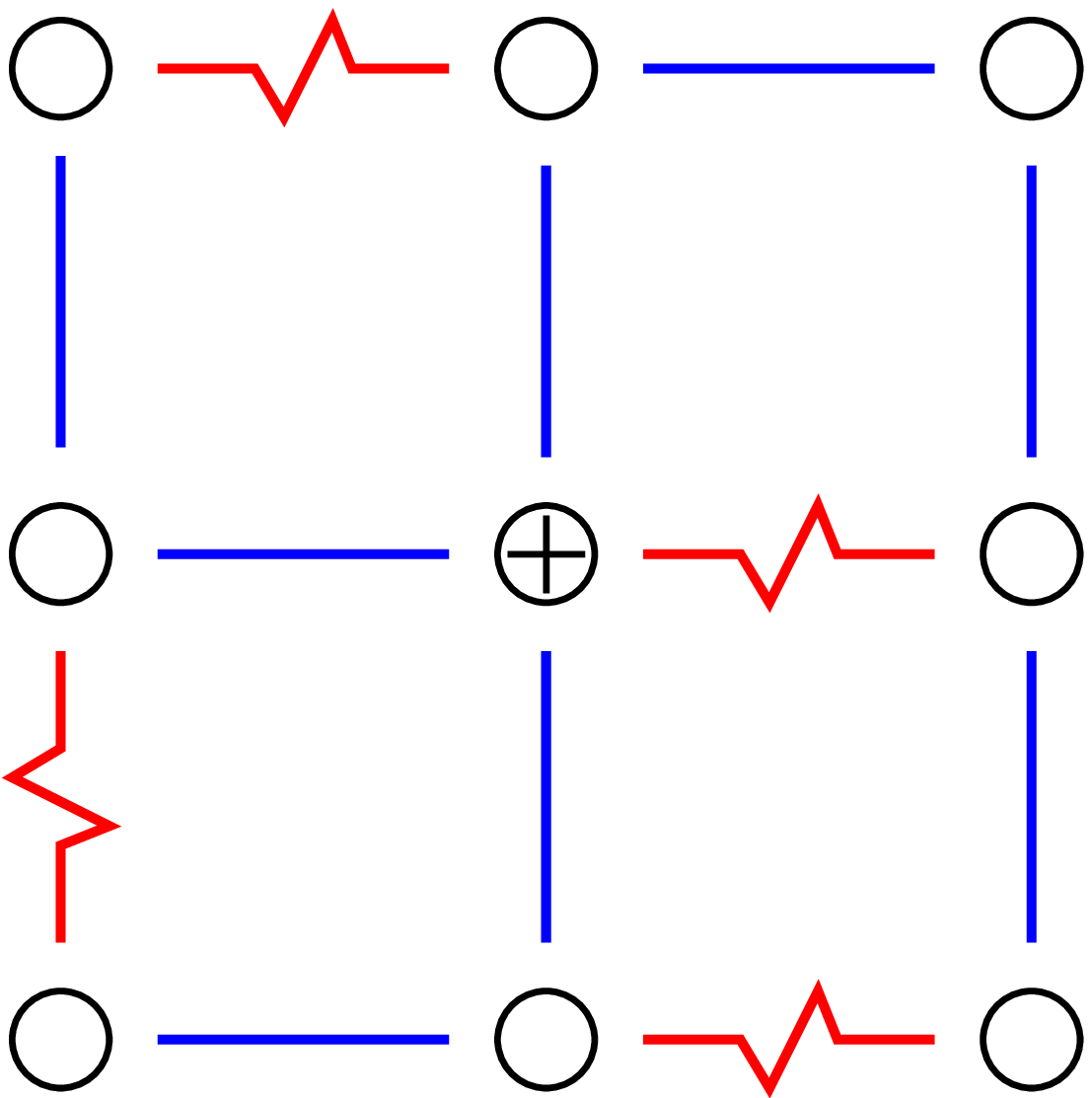}}

\vspace{0.2cm}

\myscaleboxc{\includegraphics{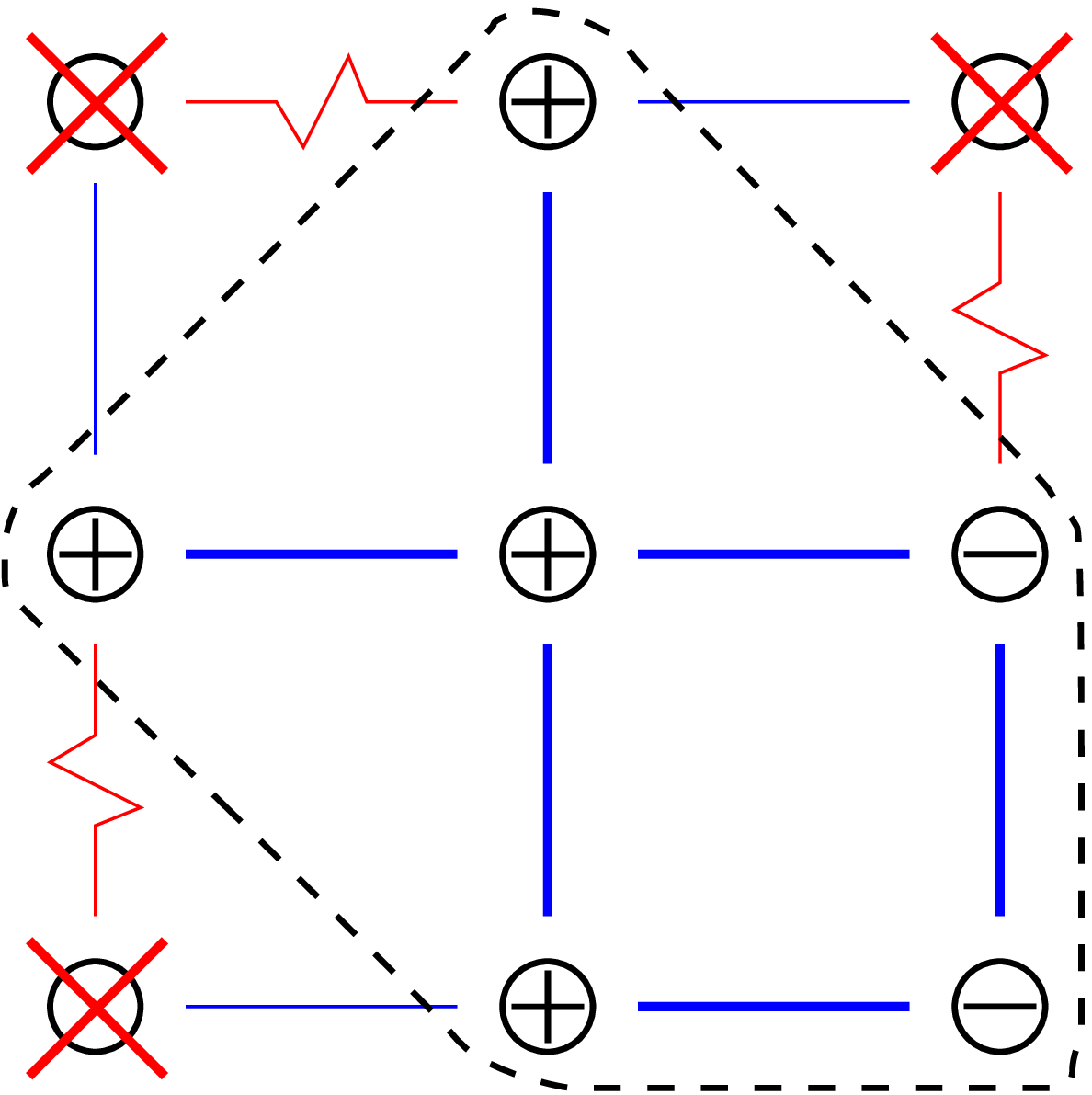}}
\end{center}
\caption{\captionCEAExample}
\label{figCEAExample}
\end{figure}


\begin{figure}[htb]
\begin{center}
\myscalebox{\includegraphics{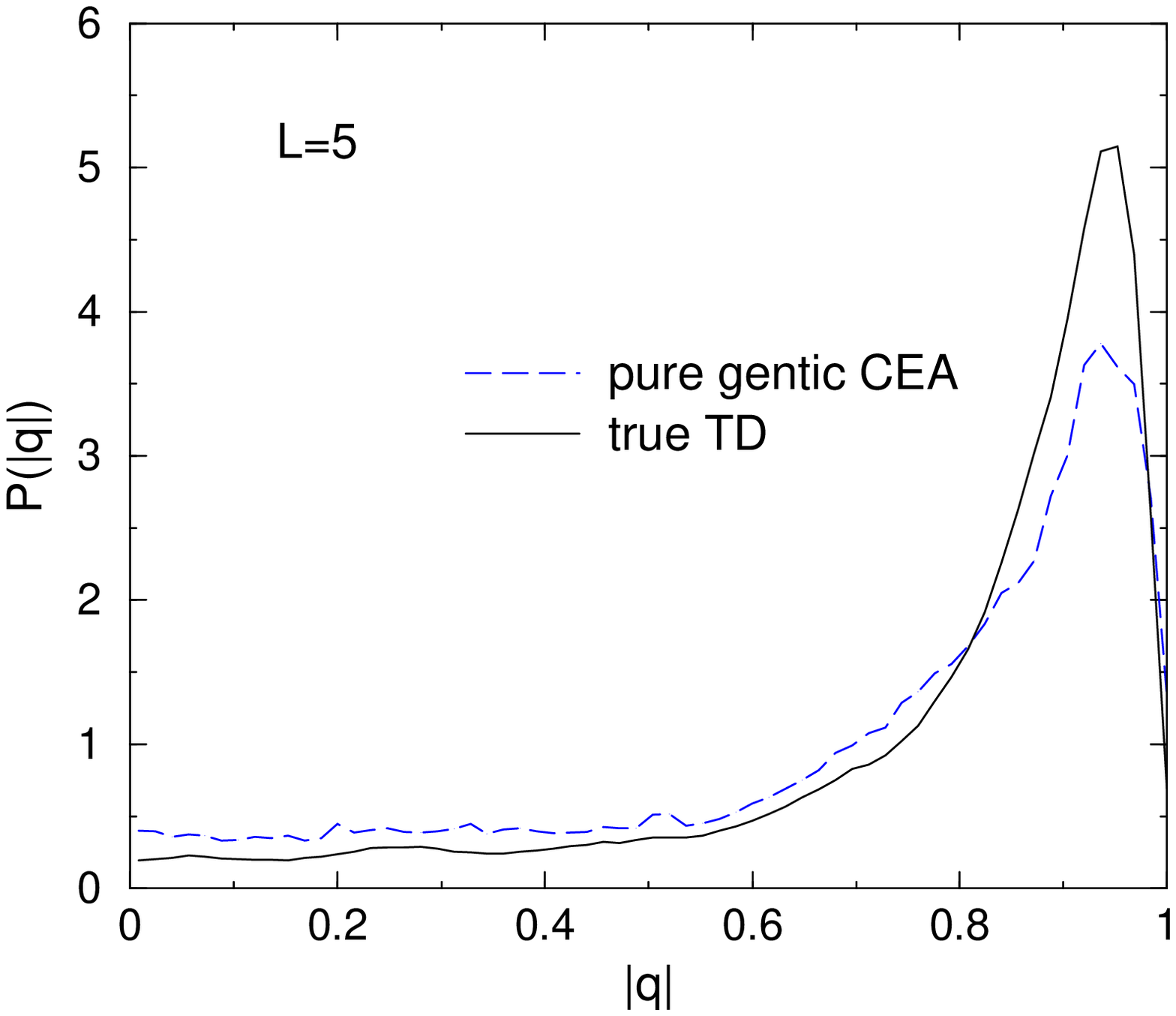}}
\end{center}
\caption{\captionPqLfive}
\label{figPqLfive}
\end{figure}


\begin{figure}[htb]
\begin{center}
\myscalebox{\includegraphics{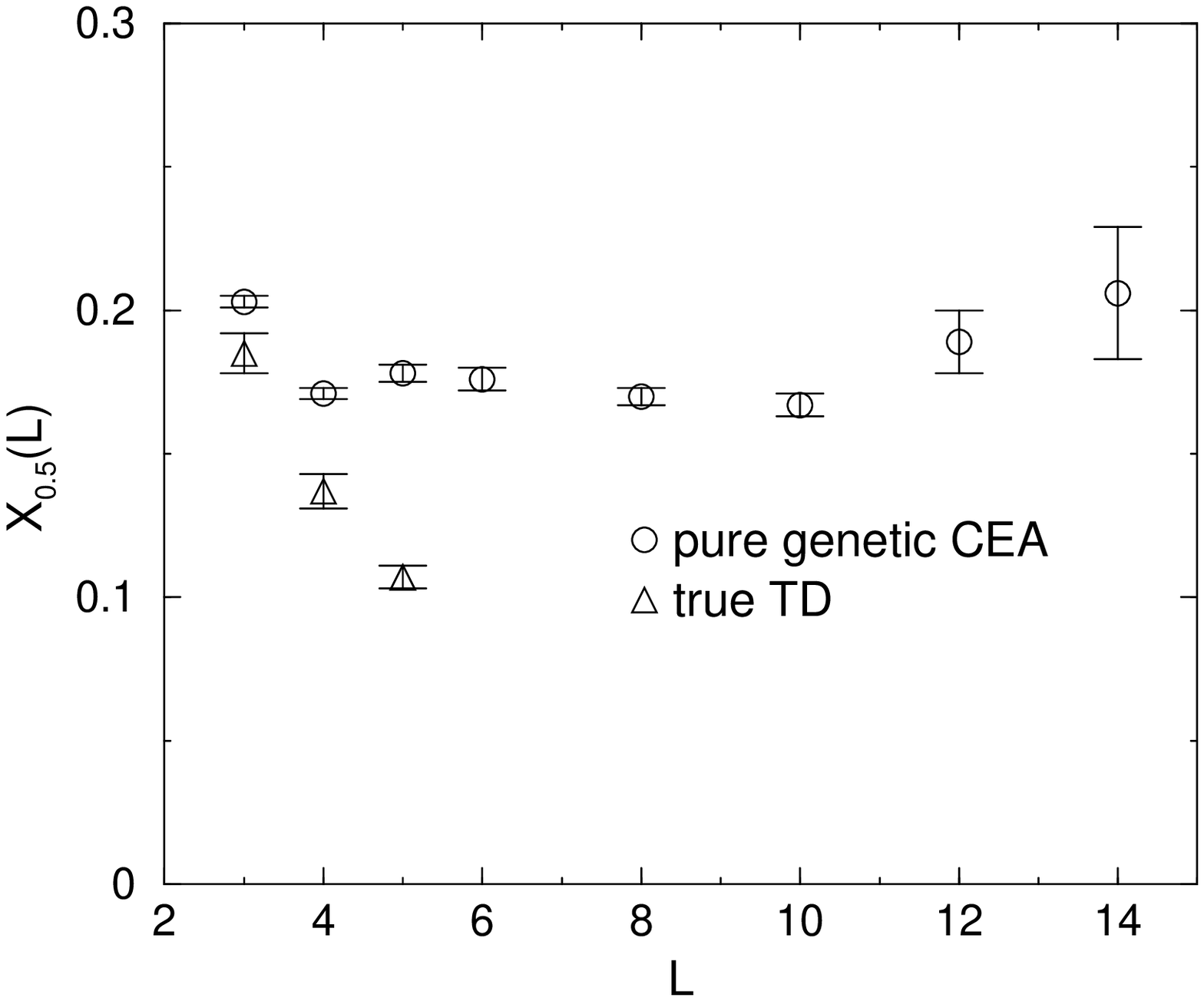}}
\end{center}
\caption{\captionXL}
\label{figXL}
\end{figure}


\begin{figure}[htb]
\begin{center}
\myscalebox{\includegraphics{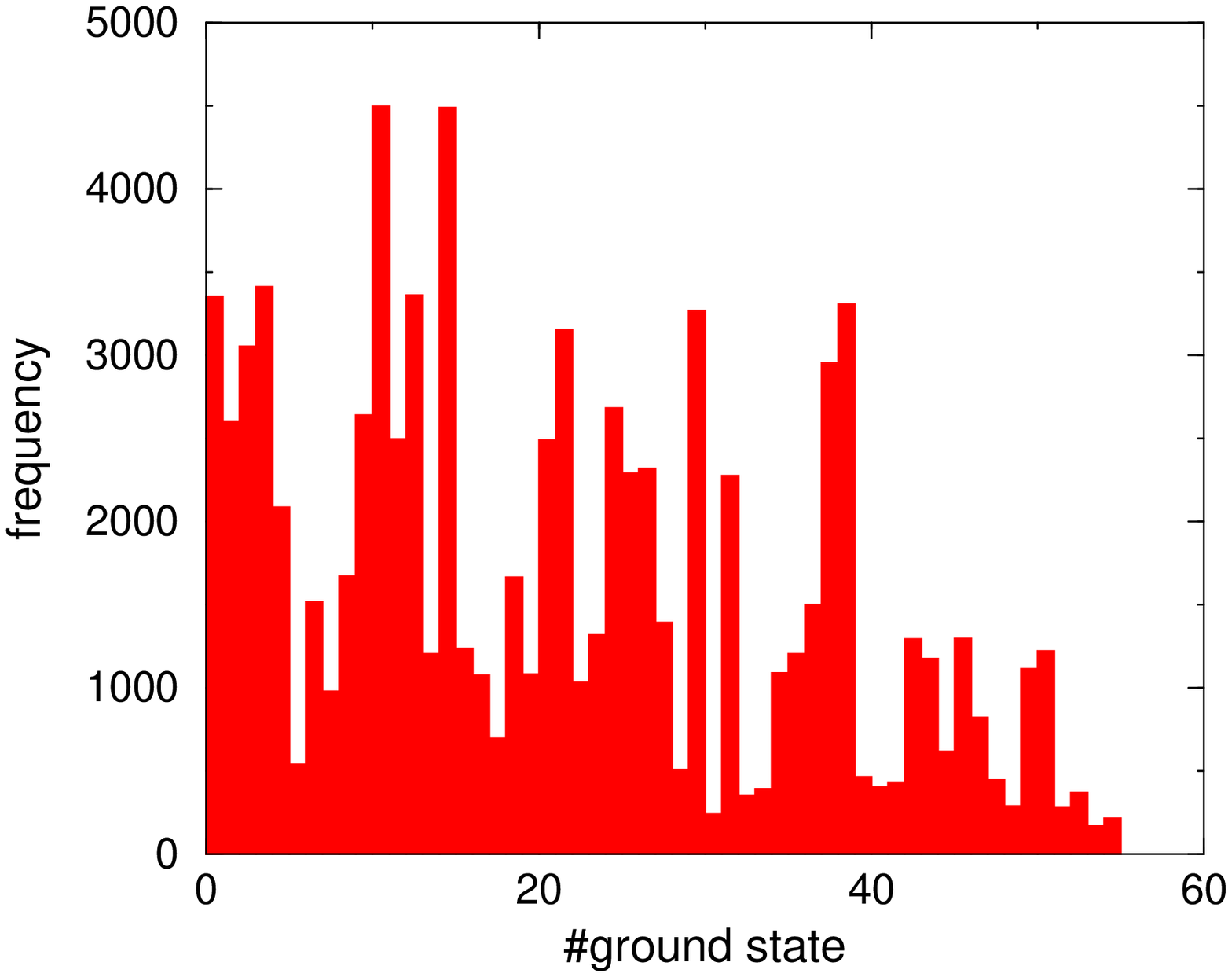}}
\end{center}
\caption{\captionHistogramm}
\label{figHistogramm}
\end{figure}


\begin{figure}[htb]
\begin{center}
\myscalebox{\includegraphics{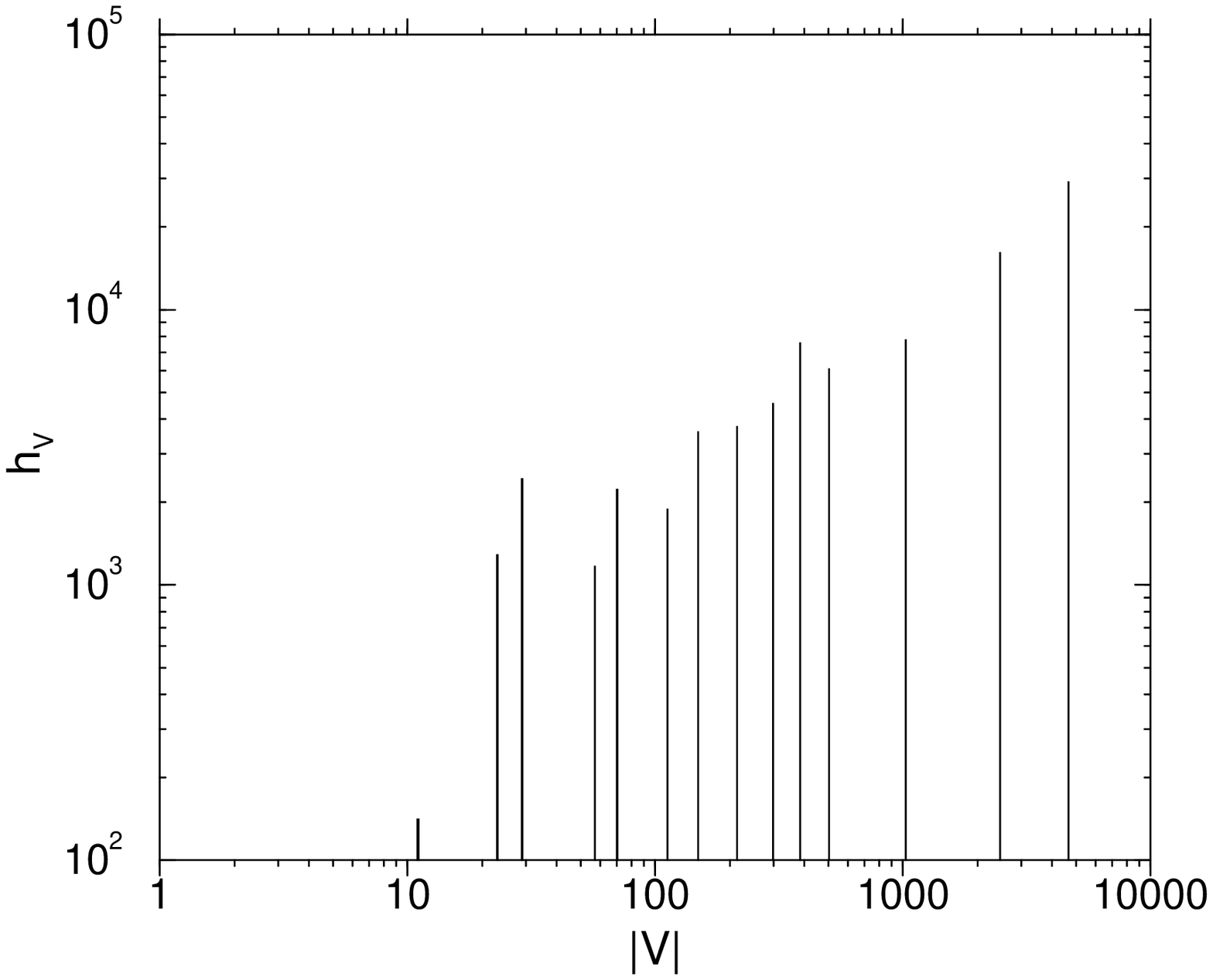}}
\end{center}
\caption{\captionHistoSample}
\label{figHistoSample}
\end{figure}


\begin{figure}[htb]
\begin{center}
\myscalebox{\includegraphics{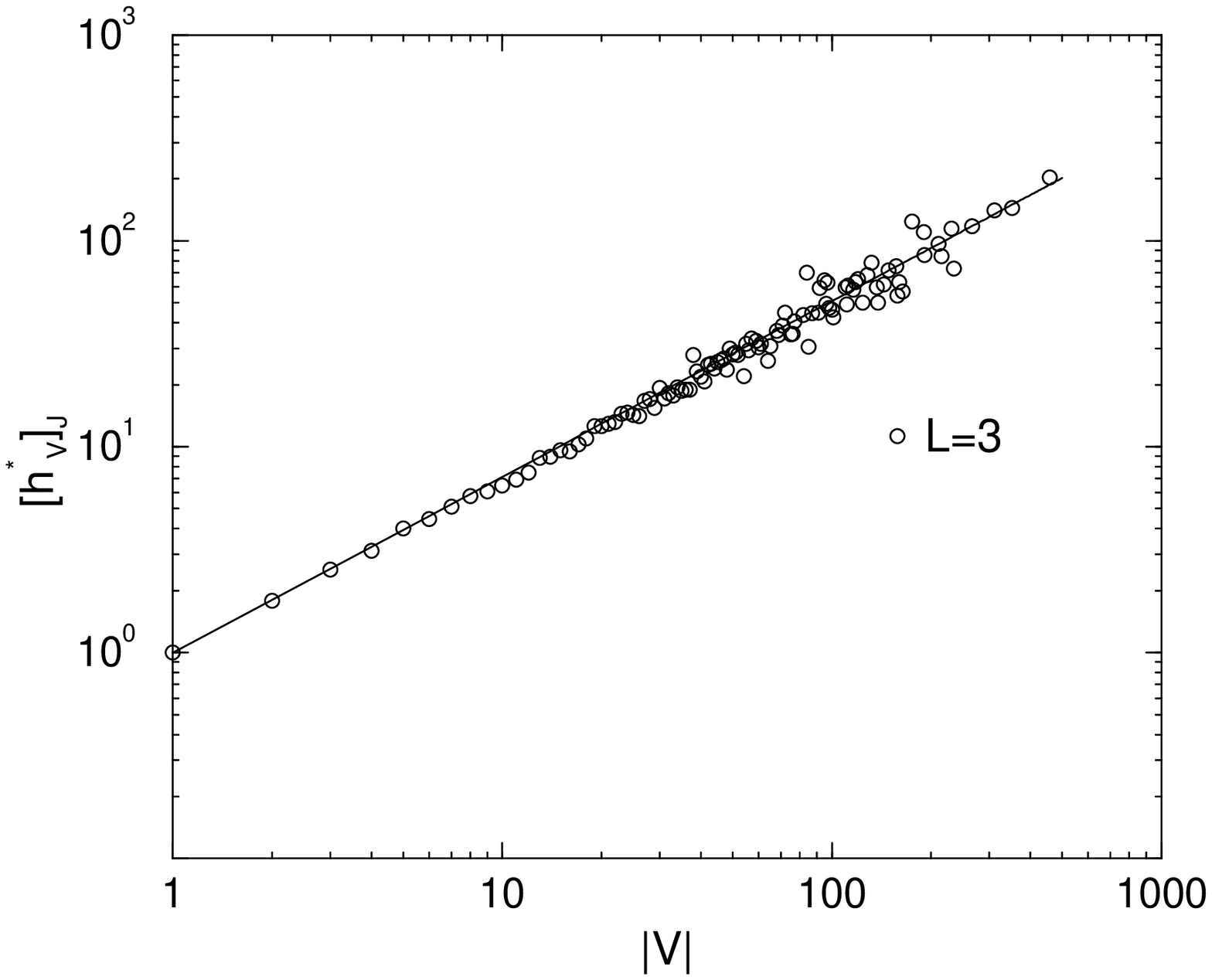}}
\end{center}
\caption{\captionHistoLthree}
\label{figHistoLthree}
\end{figure}

\begin{figure}[htb]
\begin{center}
\myscalebox{\includegraphics{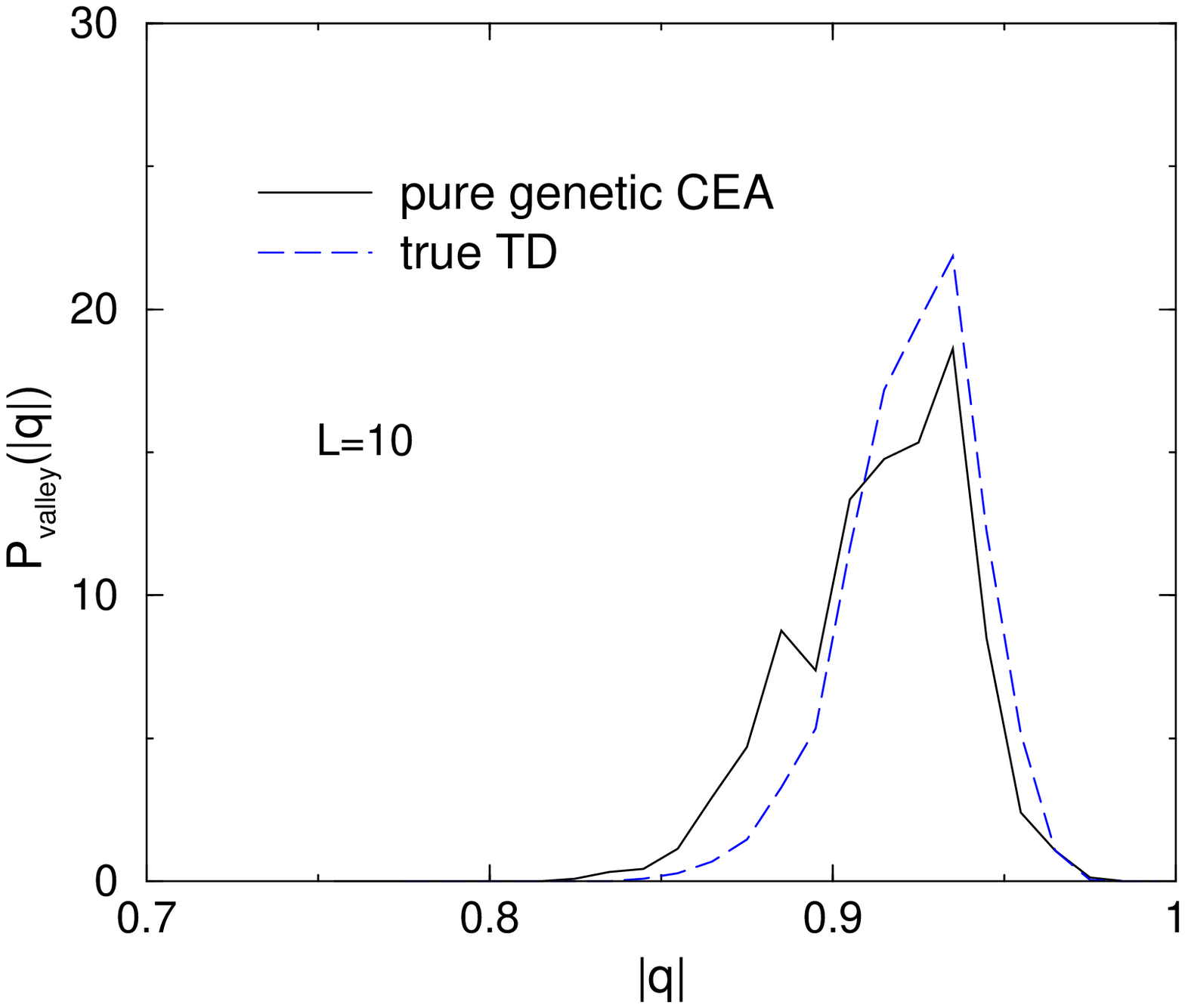}}
\end{center}
\caption{\captionPqValley}
\label{figPqValley}
\end{figure}

\end{document}